\documentclass[
 reprint,
 amsmath,amssymb,
 aps,
 noeprint,
 longbibliography,
 prl
]{revtex4-1}

\usepackage{graphicx}
\usepackage[dvipsnames]{xcolor}
\usepackage[colorlinks,linkcolor=black,citecolor=black,urlcolor=black,filecolor=black]{hyperref}
\usepackage{amsmath, amssymb, amsthm}
\usepackage{wasysym}
\usepackage{lipsum}
\usepackage{outlines}
\usepackage{enumerate}
\usepackage{setspace}
\usepackage{tabularx} 
\usepackage{xcolor}
\usepackage[export]{adjustbox}
\usepackage{tikz}
\usetikzlibrary{decorations.pathreplacing}
\usetikzlibrary{arrows,arrows.meta,calc,shapes.geometric,shapes.misc}
\tikzset{
	>=stealth',
	help lines/.style={dashed, thick},
	important line/.style={thick},
	connection/.style={thick, dotted},
}
\DeclareMathAlphabet{\mymathbb}{U}{BOONDOX-ds}{m}{n}
\usepackage{dsfont}
\usepackage{braket}
\hypersetup{
    pdfstartview={FitH}, 
    colorlinks=true, 
    linkcolor=blue, 
    citecolor=blue, 
    filecolor=magenta,
    urlcolor=blue
}
\usepackage{algorithm}
\usepackage{algpseudocode}

\usepackage{wrapfig}

\newcommand{\p}{p}
\newcommand{\pth}{p_\text{th}}
\newcommand{\fbefore}{f_b}
\newcommand{\nrounds}{n_r}
\newcommand{\ftwo}{\mathbb{F}_2}
\newcommand{\huf}{HUF}
\newcommand{\bhuf}{belief-HUF}
\newcommand{\Bhuf}{Belief-HUF}
\newcommand{\mle}{MLE}

\newcommand{\plogical}{P_{L}}
\newcommand{\plogicalmax}{P_{L, \text{max}}}
\newcommand{\plogicalround}{P_{L, \text{round}}}
\usepackage{verbatim}

\begin{document}

\title{Correlated decoding of logical algorithms with transversal gates}

\author{
Madelyn~Cain$^{1}$, Chen~Zhao$^{1,2}$, Hengyun~Zhou$^{1,2}$, Nadine~Meister$^{1}$, J. Pablo Bonilla~Ataides$^{1}$, Arthur~Jaffe$^{1}$, Dolev~Bluvstein$^{1}$, and Mikhail~D.~Lukin$^{1,*}$}
\affiliation{
$^1$Department~of~Physics,~Harvard~University,~Cambridge,~MA~02138,~USA \\ 
$^2$QuEra Computing Inc., Boston, MA 02135, USA
}

\date{\today}

\begin{abstract}
Quantum error correction is believed to be essential for scalable quantum computation, but its implementation is challenging due to its considerable space-time overhead.  
Motivated by recent experiments demonstrating efficient manipulation of logical qubits using transversal gates (Bluvstein \textit{et al.}, \href{https://doi.org/10.1038/s41586-023-06927-3}{Nature \textbf{626}, 58-65 (2024)}), we show that the performance of logical algorithms can be substantially improved by decoding the qubits jointly to account for error propagation during transversal entangling gates.
We find that such correlated decoding improves the performance of both Clifford and non-Clifford transversal entangling gates, and explore two decoders offering different computational runtimes and accuracies. 
In particular, by leveraging the deterministic propagation of stabilizer measurement errors through transversal Clifford gates, we find that correlated decoding enables the number of noisy syndrome extraction rounds between these gates to be reduced from $O(d)$ to $O(1)$ in Clifford circuits, where $d$ is the code distance. 
We verify numerically that this approach substantially reduces the space-time cost of deep logical Clifford circuits.
These results demonstrate that correlated decoding provides a major advantage in early fault-tolerant computation, as realized in recent experiments, and further indicate it has considerable potential to reduce the space-time cost in large-scale logical algorithms.
\end{abstract}

\maketitle

Quantum error correction (QEC) is believed to be essential for large-scale quantum computation~\cite{Shor1995, Preskill_1998, Dennis2002, nielsen_chuang_2010, Kitaev2003}.
It is a rapidly developing frontier, with recent milestone demonstrations of the preservation of a logical qubit and universal logic gates~\cite{Ofek2016, Reinhold2020, anderson_2021, Postler2022, ryananderson2022implementing, Andersen2023, google_2023, menendez2023implementing, Sivak2023, wang2023faulttolerant, Self2024}. 
However, implementing large-scale algorithms with protected, logical qubits is challenging due to its considerable space-time overhead.
Most recently, reconfigurable neutral atom arrays have realized quantum algorithms with tens of logical qubits and hundreds of logical entangling gates~\cite{Bluvstein2024}, opening the door to early fault-tolerant computation. 
Key to these results was the ability to efficiently perform fault-tolerant transversal entangling gates by dynamically reconfiguring qubits during the computation~\cite{Bluvstein2022}, similar to prior work with trapped ions~\cite{Postler2022, ryananderson2022implementing, menendez2023implementing, wang2023faulttolerant}. 

These experiments~\cite{Bluvstein2024} motivate theoretical explorations of the optimal ways to design and decode error-corrected circuits, in order to accurately infer which physical errors occurred such that they can be  corrected~\cite{terhal_quantum_memories, iolius2023decodingreview}.
In particular, transversal entangling gates apply the same operation to the physical qubits within the codes, resulting in highly structured error propagation: errors spread in a known way between, but not within, logical code blocks. 
As a result, errors detected on one logical qubit can contain information about which errors occurred on other logical qubits, and these correlations can be utilized to improve the accuracy of the decoder.
Such correlated decoding was already used in Ref.~\cite{Bluvstein2024}, demonstrating its utility for near-term experiments.
In addition, it may impact the resource requirements of complex large-scale quantum algorithms~\cite{Litinski_2019, Gidney2021howtofactorbit, litinski2022active}, as transversal gates can comprise core subroutines in both QEC~\cite{bravyi_kitaev_distillation, fowler_surface_code} and algorithms~\cite{cuccaro2004new}. 

In this Letter, we analyze the benefits of jointly decoding the logical qubits in an algorithm to account for error propagation during transversal entangling gates. 
We explore two decoders which utilize the space-time decoding hypergraph of the logical algorithm~\cite{gidney2021stim, gottesman2022opportunities, belief_matching_2023, McEwen2023relaxinghardware, delfosse2023spacetime, supplement}, one of which is highly accurate, and another which is approximate but guaranteed to have an efficient runtime.
We find that correlated decoding improves the performance of both Clifford and non-Clifford transversal entangling gates. 
We also show that it enables the number of syndrome extraction rounds to be substantially reduced in deep logical Clifford circuits.
These results provide the theoretical foundation for recent experimental observations~\cite{Bluvstein2024}, highlight the utility of correlated decoding in early fault-tolerant computation, and indicate its considerable potential to reduce the space-time cost of large-scale logical algorithms. 

\begin{figure*}[ht]
    \centering
    \includegraphics{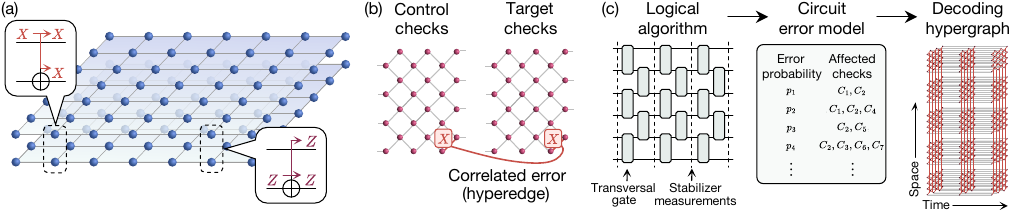}
    \caption{Decoding logical algorithms with transversal gates. 
    (a)~Transversal CNOT gates copy physical errors between logical qubits. 
    (b)~
    These transferred errors flip the same stabilizers on both logical qubits after the CNOT, generating a hyperedge in the decoding hypergraph connecting the control and target check vertices.
    (c)~Given a physical error model, the decoding hypergraph for a logical algorithm can track these transferred errors, which can appear as hyperedges connecting checks from multiple logical qubits at different points in time.
    }
    \label{fig:figure_1}
\end{figure*}

The key idea of this work can be understood as follows. 
We consider a quantum algorithm with logical qubits, each comprised of redundant physical qubits~\cite{gottesman2009introduction, Dennis2002} (see Fig.~\ref{fig:figure_1}(a)).
The logical information of each code is stored in the simultaneous $+1$ eigenspace of its stabilizer operators. 
Physical errors can be detected by measuring the stabilizers, as any error which anticommutes with a stabilizer will flip its measurement outcome to $-1$.
A decoding algorithm can use these stabilizer measurements, or syndromes, to predict which physical error(s) likely caused the observed syndromes. 
Decoding succeeds if the associated correction does not erroneously apply a logical operation. 

Errors during transversal entangling gates can create correlations in the stabilizer measurements of different logical qubits, which can be utilized to improve the accuracy of the decoder. 
In a transversal CNOT, for example, physical $X$ errors are copied from the control to the target qubit, and $Z$ errors are copied from the target to the control qubit~(Fig.~\ref{fig:figure_1}(a)).
Although this propagation increases the density of errors, it is deterministic and can be deduced from correlations in the measured stabilizers. 
For instance, assuming that $X$ errors occur only before a transversal CNOT, one can infer that a flipped $Z$ stabilizer on both logical qubits originates from a single $X$ error copied from the control, effectively halving the density of errors in the decoding. 
Similarly, a physical error can propagate onto $N$ logical qubits after $N$ sequential CNOTs.
Although this process flips many stabilizers, tracking the error propagation again reveals that the syndrome originated from just a single error.

\textit{Correlated decoding algorithms.---} 
To formalize this approach, we define the decoding hypergraph of the logical algorithm~\cite{fowler2010surfacecodequantumerror, gidney2021stim, gottesman2022opportunities, belief_matching_2023, McEwen2023relaxinghardware, delfosse2023spacetime}, which relates each possible physical error mechanism to the stabilizer measurement(s) it flips. 
The hypergraph vertices correspond to $N$ measured \textit{checks} of the logical qubits $\{C_i\}_{i=1\dots N}$, which compare consecutive stabilizer measurements in time.
Each check is defined as the product of a stabilizer measurement and its previous measurement (if one exists), obtained from back-propagating the stabilizer operator through the circuit to the point it was previously measured.
If one stabilizer measurement in the check detects an error, the check flips from $+1$ to $-1$.
Each error mechanism is associated with a hyperedge connecting the check(s) it flips. 
Errors copied during transversal entangling gates therefore correspond to hyperedges connecting the checks of multiple logical qubits~(Fig.~\ref{fig:figure_1}(b)).
We use Stim~\cite{gidney2021stim}, a Clifford circuit simulator, to identify the $M$ hyperedges $\{E_j\}_{j=1\dots M}$ and their probabilities $\{p_j\}_{j=1\dots M}$ under a chosen circuit error model, and which error source(s) $I(C_i)$ flip each check $C_i$~(Fig.~\ref{fig:figure_1}(c)).

In order to realize QEC in practice, a specific decoder should be chosen. 
Many conventional decoders such as minimum weight perfect matching (MWPM)~\cite{higgott2021pymatchingpython, higgott2023sparse}, a widely used surface code decoder, are constrained to operate on decoding \textit{graphs}, whose edges connect at most two vertices. 
In contrast, here we explore two decoders which take as input a general decoding hypergraph of a logical algorithm, which can include hyperedges connecting more than two vertices. 
The first decoder exactly computes the most-likely error (\mle{}) consistent with the measured syndrome. 
It maximizes the total error probability \mbox{$\prod_{j=1}^M p_j^{E_j} (1-p_j)^{1-E_j}$} over the binary variables $E_j$, which are equal to one if the error $E_j$ occurred and zero otherwise. 
The errors are constrained to be consistent with the measured syndrome: if $C_i = +1$ $(-1)$, then an even (odd) number of errors occurred in $I(C_i)$. 
We map this problem to a mixed-integer program, and find its optimal solution using a state-of-the-art solver~\cite{landahl2011faulttolerant, takada2023highly, gurobi, supplement}.

Because finding the \mle{} exactly is NP-hard, we expect the runtime of the above algorithm to grow exponentially in $M$ in the worst-case~\cite{berlekamp_np_hard_decoding_1978, vardy_1997}.
This motivates our second algorithm, hypergraph union-find with belief propagation (\bhuf{})~\cite{Delfosse2021almostlineartime, delfosse2021unionfind}, which instead approximates the \mle{} and has an efficient worst-case time complexity of $O(M^3\log(M))$~\cite{supplement}.
\Bhuf{} first uses belief propagation to update the error probabilities $\{p_j\}_{j=1\dots M}$ to approximate the probability of each error given the measured syndrome~\cite{belief_matching_2023}.
Leveraging the insight that errors are typically localized, \bhuf{} then expands clusters of error mechanisms and checks on the decoding hypergraph at a rate dependent on the error probabilities, merging clusters that meet.
A cluster stops expanding when it contains an error mechanism consistent with its measured checks, and the algorithm terminates when all clusters stop expanding.

\textit{Decoding transversal Clifford gates.---}
As a simple illustrative example, we apply these techniques to decode a transversal CNOT, which is available for all Calderbank-Shor-Steane (CSS) codes~\cite{shor1997faulttolerant}. 
A key property of error-correcting circuits is that their performance can be improved for physical error rates below some threshold $\pth$ by increasing the code distance (the minimum weight of a logical Pauli operator), which is related to how many errors can be corrected~\cite{Dennis2002, fowler_surface_code}. 
In practice, the threshold depends on the decoder, as higher-accuracy decoders can correct more complex patterns of errors.
To explore the optimal advantage of correlated decoding, we will compare decoding two surface codes jointly with \mle{}, and independently with MWPM~\cite{higgott2021pymatchingpython, higgott2023sparse}.

\begin{figure}[t!]
    \centering
    \includegraphics{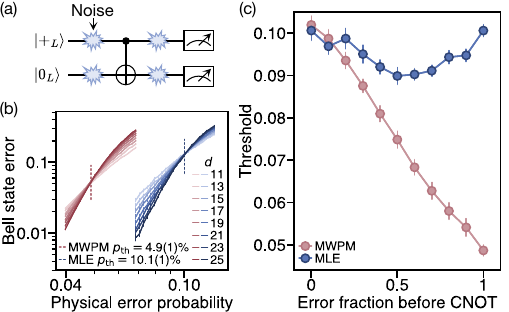}
    \caption{Decoding the transversal CNOT. 
    (a)~
    We generate a logical Bell pair using a noisy transversal CNOT, with physical errors before and after the CNOT with probabilities $\p\fbefore$ and \mbox{$\p(1-\fbefore)$}, respectively.
    (b)~When the errors occur before the CNOT ($\fbefore = 1$), the threshold of uncorrelated MWPM (pink) is half that of \mle{} (blue).
    (c)~As $\fbefore$ increases, more errors are transferred between the logical qubits, and \mle{} increasingly outperforms MWPM.
    }
    \label{fig:figure_2}
\end{figure}

\begin{figure*}[t]
    \centering
    \includegraphics{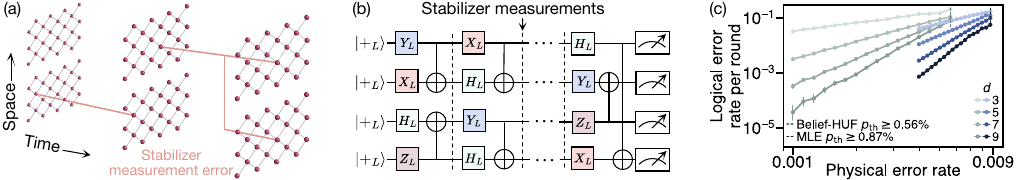}
    \caption{Correlated decoding of deep logical circuits. 
    (a)~Stabilizer measurement errors near transversal CNOTs can appear as hyperedges connecting the checks of both logical qubits at different times. 
    (b)~We study deep logical Clifford circuits with $\nrounds$ rounds of syndrome extraction between CNOTs. 
    (c)~Setting $\nrounds = 1$, our results are consistent with thresholds of $\pth \geq 0.56\%$ for \bhuf{} (green) and $\pth \geq 0.87\%$ for \mle{} (blue). 
    }
    \label{fig:figure_3}
\end{figure*}

We study the performance of the transversal CNOT by numerically generating a Bell pair between rotated surface codes, followed by noiseless syndrome extraction. 
Using Stim~\cite{gidney2021stim}, we prepare the codes noiselessly in $\ket{+_L 0_L}$, then perform a CNOT with single-physical-qubit $X$ and $Z$ errors before and after with probabilities $p\fbefore$ and \mbox{$p(1-\fbefore)$}, respectively~(Fig.~\ref{fig:figure_2}(a)). 
We use this simplified error model, parameterized by $\fbefore$, to modulate the fraction of errors transferred between qubits during the CNOT.
Following the CNOT, the codes are first measured noiselessly in the $X$ or $Z$ basis to extract the syndrome, then decoded to compute the mean of the $X^1_LX^2_L$ and $Z^1_LZ^2_L$ populations as a proxy for fidelity~\cite{entanglement_witness}. 
Our decoding hypergraph includes vertices corresponding to the final stabilizer measurements, and inter- and intra-logical qubit hyperedges corresponding to the possible errors before and after the CNOT~(see Fig.~\ref{fig:figure_1}(b))~\cite{supplement}.

Because errors before the CNOT are copied between logical qubits, we expect the benefits of correlated decoding to grow with $\fbefore$. 
When the errors occur exclusively before the CNOT \mbox{($\fbefore = 1$)}, the density of errors on one logical qubit is doubled due to these copied errors. 
Therefore, the numerically-fitted~\cite{supplement} threshold of MWPM \mbox{($\pth = 4.9(1)\%$)} is half that of \mle{} \mbox{($\pth = 10.1(1)\%$)} (Fig.~\ref{fig:figure_2}(b)). 
In Figure~\ref{fig:figure_2}(c), we study the thresholds of \mle{} and MWPM as a function of $\fbefore$. 
When $\fbefore = 0$, no errors are transferred between logical qubits, so the thresholds of MWPM and \mle{} are equal. 

\textit{Decoding deep logical Clifford circuits.---}
In realistic quantum algorithms, stabilizer measurements are used to remove entropy from the logical qubit by tracking physical errors over time~\cite{gottesman2009introduction}. 
A challenge, however, is that the stabilizer measurements are also noisy. 
As a result, entangling gates based on lattice surgery~\cite{horsman2012surfacecode} and standard error correction constructions~\cite{shor1997faulttolerant, gottesman2009introduction} require $d$ rounds of syndrome extraction to ensure fault tolerance to these errors, creating considerable space-time overhead. 
We now explore if we can perform fewer than $d$ rounds between transversal CNOTs, by leveraging the deterministic propagation of stabilizer measurement errors through CNOTs to verify stabilizer measurements using surrounding rounds of syndrome extraction. 

When constructing the decoding hypergraphs of circuits with noisy syndrome extraction, however, we observe that stabilizer measurement errors near transversal CNOTs generate hyperedges that make conventional decoding approaches challenging. 
To see this, consider two surface codes with syndrome extraction at times $t-1, t,$ and $t+1$, with a transversal CNOT controlled on qubit 1 between times $t$ and $t+1$. 
The checks associated with a particular $Z$ stabilizer on both qubits are $\mathcal{Z}_{t-1}^1 \mathcal{Z}_{t}^1$ and $\mathcal{Z}_{t-1}^2 \mathcal{Z}_{t}^2$ ($\mathcal{Z}_{t}^1 \mathcal{Z}_{t+1}^1$ and $\mathcal{Z}_{t}^2 \mathcal{Z}_{t}^1 \mathcal{Z}_{t+1}^2$) for the first (last) two rounds, the latter of which comes from back-propagating the stabilizers through the CNOT.
A measurement error on $\mathcal{Z}_{t}^1$ will thus flip three checks: $\mathcal{Z}_{t-1}^1 \mathcal{Z}_{t}^1$, $\mathcal{Z}_{t}^1 \mathcal{Z}_{t+1}^1$, and $\mathcal{Z}_{t}^2 \mathcal{Z}_{t}^1 \mathcal{Z}_{t+1}^2$~(Fig.~\ref{fig:figure_3}(a)).
Because this hyperedge has order three (meaning it connects three vertices) and cannot be decomposed into existing lower-order edges~\cite{higgott2021pymatchingpython, supplement}, MWPM cannot be directly applied, as it only operates on edges of order $\leq 2$.
In the Supplementary Materials~\cite{supplement}, we find that MWPM can be applied at the expense of $d$ syndrome extraction rounds before a CNOT, but its threshold is substantially lower than that of the correlated decoders.

In contrast, because the correlated decoders can be applied to general decoding hypergraphs, they can potentially decode circuits with fewer than $d$ syndrome extraction rounds between CNOTs. 
To study this, we numerically simulate a deep logical Clifford circuit between four surface codes. 
The codes are initialized noiselessly in $\ket{+_L}^{\otimes 4}$, then entangled with 32 layers of transversal gates, with $\nrounds$ rounds of syndrome extraction between layers~(Fig.~\ref{fig:figure_3}(b)). 
Each layer is comprised of a random transversal Hadamard $H_L$ or Pauli gate $\{X_L, Y_L, Z_L\}$, followed by a random pairing of CNOTs with randomized designation of control and target.
We add depolarizing noise to each circuit-level physical operation with probability $p$, as in Refs.~\cite{belief_matching_2023, supplement}. 
Finally, we noiselessly measure the $X$ and $Z$ stabilizers and the logical stabilizers of the resulting state, then decode to find the logical error rate $\plogical$~\cite{gottesman1998heisenberg}.
In the Supplementary Materials~\cite{supplement}, we confirm that our results are insensitive to boundary effects from the final noiseless measurements, and representative of different randomly sampled circuits. 

\begin{figure}[b]
    \centering
    \includegraphics{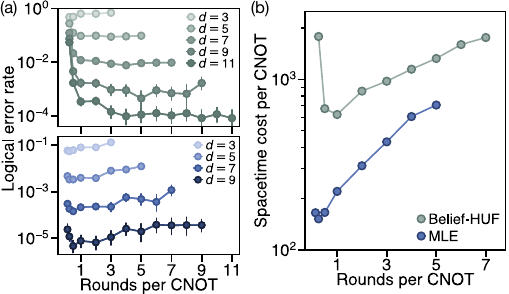}
    \caption{Reducing the space-time cost of deep logical circuits. 
    (a)~The rounds of syndrome extraction per CNOT can be reduced to $\nrounds\simeq \frac{1}{2}$ for \mle{} (bottom) and $\nrounds\simeq 3$ for \bhuf{} (top), without increasing $P_L$. 
    (b)~The extrapolated space-time cost to reach $\plogical = 10^{-6}$ is minimized at $\nrounds = 1$ for \bhuf{} and $\nrounds = \frac{1}{4}$ for \mle{}.}
    \label{fig:figure_4}
\end{figure}

We first study the case of one round of syndrome extraction per CNOT (\mbox{$\nrounds=1$}). 
Figure~\ref{fig:figure_3}(c) shows the logical error rate per round, \mbox{$\plogicalmax\big[1 - (1 - \plogical/\plogicalmax)^{1/(32\nrounds)}\big]$}, where \mbox{$\plogicalmax = 1-\frac{1}{2^4}$} is the error rate of the maximally mixed logical state, as a function of $p$~\cite{supplement}. 
The numerical results are consistent with the existence of a threshold for both \bhuf{} and \mle{}. 
Though we cannot reliably fit the thresholds because \mbox{$\plogical/\plogicalmax \to 1$} near the threshold, we estimate that $\pth \geq 0.56\%$ for \bhuf{} and $\pth \geq 0.87\%$ for \mle{} from the largest physical error rate at which $\plogical$ decreases with $d$~\cite{supplement}. 
Furthermore, in the Supplementary Materials~\cite{supplement} we observe that the MLE logical error rate scaling is exponentially suppressed in $\lfloor \frac{d+1}{2}\rfloor$. 

Next, we investigate whether the space-time cost per CNOT, at a fixed physical error rate of $p=0.001$, can be reduced by optimizing $\nrounds$. 
We express the space-time cost as $(\nrounds + 1) d^2$, corresponding to a cost of $d^2$ for the CNOT and $\nrounds d^2$ for syndrome extraction.
In Figure~\ref{fig:figure_4}(a), we plot $P_L$ for various code distances and $\nrounds\in\{\frac{1}{8}, \frac{1}{4}, \frac{1}{2}, 1, \dots, d\}$ ($\nrounds < 1$ corresponds to multiple CNOTs before a round of syndrome extraction).
These results suggest that $\nrounds$ can be reduced to $\nrounds \simeq \frac{1}{2}$ and $\nrounds \simeq 3$ for \mle{} and \bhuf{}, respectively, without increasing $\plogical$. 
$P_L$ is expected to decrease exponentially with $d$ according to the heuristic formula~\cite{fowler_surface_code, supplement}
\begin{equation}
    \plogical(\nrounds, d) = 32 C \nrounds (\alpha_{\nrounds})^{\frac{d+1}{2}}.
\end{equation}
By fitting the coefficient $C$ and exponent base $\alpha_{\nrounds}$, we extrapolate the code distance  needed to suppress $\plogical$ to $10^{-6}$, from which we compute the space-time cost per CNOT~
(Fig.~\ref{fig:figure_4}(b)). 
The space-time cost is minimized at $\nrounds = 1$ for \bhuf{} and $\nrounds = \frac{1}{4}$ for \mle{}.
Note that at $\nrounds = \frac{1}{4}$, the errors from the four transversal CNOTs balance the four physical CNOT layers from syndrome extraction~\cite{fowler_surface_code, supplement}.
Therefore, the space-time cost of these logical circuits can be substantially reduced by using correlated decoding to optimize $\nrounds$. 

\begin{figure}[h]
    \centering
    \includegraphics{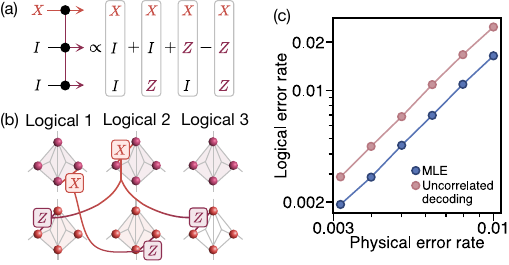}
    \caption{Decoding the transversal CCZ. 
    (a)~A physical $XII$ error before a CCZ propagates to \mbox{$\frac{1}{2}(XII + XIZ + XZI - XZZ)$}. 
    (b)~The decoding hypergraph includes a hyperedge for each Pauli error in the superposition.
    (c)~The logical error rate is suppressed by a factor of $\simeq 1.5$ by decoding the logical qubits jointly. }
    \label{fig:figure_5}
\end{figure}

\textit{Decoding transversal non-Clifford gates.---}
Correlated decoding can also be applied to non-Clifford transversal gates. 
As a proof-of-concept, we study a transversal CCZ between three fifteen-qubit quantum Hamming codes, with perfect syndrome extraction~\cite{paetznick_ccz_2013, supplement}. 
We perfectly initialize each code in an $X$ or $Z$-basis product state, then perform a CCZ with single-physical-qubit depolarizing noise beforehand. 
Because the CCZ is non-Clifford, these errors can propagate to a superposition of Pauli errors on multiple logical qubits, each generating a hyperedge in the decoding hypergraph~(Fig.~\ref{fig:figure_5}(a-b))~\cite{supplement, bombin2018transversalgateserrorpropagation, scruby_nonpauli_2022}.
We analyze initial states for which the circuit outputs a logical stabilizer state, which can be efficiently simulated~\cite{supplement}. 
We then noiselessly measure the $X$ and $Z$ stabilizers and logical stabilizers, and decode to compute the average logical error rate over initial states~\cite{gottesman1998heisenberg}.
Figure~\ref{fig:figure_5}(c) shows that correlated decoding with \mle{} outperforms decoding the qubits independently by a factor of $\simeq 1.5$. 
Note that the gain here is smaller than with Clifford gates, as the error propagation is no longer fully deterministic. 
This motivates additional analysis of correlated decoding with non-Clifford gates as an avenue for future research.

\textit{Outlook.---}
Our results show that correlated decoding constitutes a powerful tool for improving the performance of logical algorithms.  
These insights were used in the experiments of Ref.~\cite{Bluvstein2024} to entangle a logical Bell pair between surface codes and to fault-tolerantly prepare logical GHZ states.
The present work demonstrates that because physical and stabilizer measurement errors can be tracked, these same techniques allow reducing the number of syndrome extraction rounds from $O(d)$ to $O(1)$ in transversal Clifford circuits.

Our correlated decoding technique can be extended in several directions. 
Although the computational runtime of \bhuf{} is efficient, its threshold is lower than that of \mle{} on certain circuits~(Fig.~\ref{fig:figure_3}(c)). 
Alternative decoders, such as those utilizing tensor networks~\cite{bravyi2014mldsurface,ferris2014tensornetworks,chubb2021generaltensor}, belief propagation with ordered statistics decoding~\cite{Panteleev2021degeneratequantum}, or machine learning~\cite{neural_decoder_2017, nn_bp_2019, nn_comparison_2020, Kuo2022, bausch2023learning, cao2023qecgpt, maan2024machinelearningmessagepassingscalable, wang2023transformerqec, ninkovic2024decodingquantumldpccodes} could potentially close this performance gap.
The runtime of \bhuf{} can be further optimized by iteratively decoding fault-tolerant blocks rather than the full circuit~\cite{bombin2023modulardecoding, skoric2023parallelwindow, alibaba_window_2023}.
Moreover, the decoders we consider are readily applicable to other codes and gates, such as the recently developed high-rate quantum LDPC codes~\cite{tillich_2014, Panteleev2021degeneratequantum, leverrier2022quantumtannercodes, pantaleev_linear_distance, pantaleev_kalachev,  breuckmann_ldpc, xu2023constantoverhead}, space-time codes~\cite{gottesman2022opportunities, delfosse2023spacetime}, and the transversal $S$ gate in the two-dimensional color code~\cite{efficient_decoding_color, projection_decoder_color, gidney2023newcircuits}.
These techniques can also be adapted to realistic experimental noise models, such as non-Pauli errors, biased noise~\cite{cong_2022}, and erasure errors~\cite{Wu2022, Ma2023, sahay_erasure_biased}. 

Another important question involves the extent to which the present techniques can be used to achieve practical speedups in classically complex circuits involving non-Clifford operations and feed-forward. 
In a recent complementary manuscript, we show that the methods described in this work can indeed be utilized for an $O(d)$ time reduction for such {\it universal} computation~\cite{zhou2024algorithmicfaulttolerancefast}.
Taken together, these results demonstrate that correlated decoding is emerging as a core building block both in new theories of fault-tolerance and in practical reductions to the cost of large-scale computation~\cite{zhou2024algorithmicfaulttolerancefast}. 
Improving the decoders and extending this technique to new applications are thus key next steps.
\\

\newpage
\let\oldaddcontentsline\addcontentsline
\renewcommand{\addcontentsline}[3]{}
\begin{acknowledgments}
\textit{Acknowledgements.---}
We thank C. Duckering for helpful suggestions on implementing the \bhuf{} algorithm. 
We also thank S.~Ebadi, S.~Evered, S.~Geim, A.~Gu, M.~Gullans, M.~Kalinowski, A.~Kubica, S.~Li, T.~Manovitz, N.~Maskara, and C.~Pattison for helpful discussions.
We acknowledge financial support from IARPA and the Army Research Office, under the Entangled Logical Qubits program (Cooperative Agreement Number W911NF-23-2-0219), the DARPA ONISQ program (grant number W911NF2010021), the DARPA IMPAQT program (grant number HR0011-23-3-0012), the Center for Ultracold Atoms (a NSF Physics Frontiers Center, PHY-1734011), the National Science Foundation (grant number PHY-2012023 and grant number CCF-2313084), the NSF EAGER program (grant number CHE-2037687), the Army Research Office MURI (grant number W911NF-20-1-0082), the Army Research Office (award number W911NF2320219 and grant number W911NF-19-1-0302), and QuEra Computing.
M.C. acknowledges support from Department of Energy Computational Science Graduate Fellowship under award number DE-SC0020347.  
J.P.B.A. acknowledges support from the Generation Q G2 fellowship and the Ramsay Centre for Western Civilisation.
D.B. acknowledges support from the NSF Graduate Research Fellowship Program (grant DGE1745303) and The Fannie and John Hertz Foundation.
This research was developed with funding from the Defense Advanced Research Projects Agency (DARPA). 
The views, opinions, and/or findings expressed are those of the author(s) and should not be interpreted as representing the official views or policies of the Department of Defense or the U.S. Government.
\end{acknowledgments}

\bibliographystyle{apsrev4-2}
\bibliography{correlated_decoding.bbl}

\clearpage
\onecolumngrid
\subsection*{\Large Supplementary Materials}
\normalsize

\let\addcontentsline\oldaddcontentsline

\setcounter{equation}{0}
\setcounter{figure}{0}
\setcounter{table}{0}
\setcounter{page}{8}
\makeatletter
\renewcommand{\theequation}{S\arabic{equation}}
\renewcommand{\thefigure}{S\arabic{figure}}
\renewcommand{\thetable}{S\arabic{table}}
\renewcommand{\refname}{}

\tableofcontents

\newpage

\section{Decoding algorithms \label{section:decoding_algorithms}}
Here we describe the implementation details of the two decoding algorithms used to decode the transversal logical circuits considered in this work. 
Both decoders take in the check measurements and the decoding hypergraph, and return a physical error which could have generated the measured syndrome. 

\subsection{MLE decoder \label{subsection:mle}}
The goal of the most-likely error (\mle{}) decoder is to exactly solve for the physical MLE consistent with the measured syndrome. 
We map the problem of finding the MLE onto a mixed-integer program, similar to the approaches in Refs.~\cite{landahl2011faulttolerant, takada2023highly}. 
Mixed-integer programs maximize a linear objective function over integer, binary, or real variables, subject to linear constraints on the variables. 
To solve for the MLE, we associate each potential physical error mechanism in the circuit (hyperedge in the decoding hypergraph) with a binary variable $E_j$ that is equal to one if that error occurs, and zero otherwise. 
Our objective is then to maximize the total error probability, \mbox{$\prod_{j=1}^M p_j^{E_j} (1-p_j)^{1-E_j}$}, over the error variables $\{E_j\}_{j=1 \dots M}$. 
To make this objective function linear in the error variables, we will maximize the logarithm of the total error probability, 
\begin{equation}
    \log\Bigg(\prod_{j=1}^M p_j^{E_j} (1-p_j)^{1-E_j}\Bigg) = \sum_{j=1}^M \log(p_j) E_j + \log(1-p_j)(1-E_j).
\end{equation}

The total error is then constrained to be consistent with the measured syndrome. 
This means that if the $i$th check $C_i = +1$ $(-1)$, then an even (odd) number of errors occurred in the set of errors which flip $C_i$, denoted $I(C_i)$. 
This constraint can be stated equivalently as \mbox{$\sum_{E_j\in I(C_i)} E_j = \frac{1}{2}(1+C_i) \mod 2$}, which is almost linear, except for the modulo. 
To linearize the constraint, we introduce an integer slack variable $K_i$ to the $i$th constraint which can add any multiple of $2$ to the summation.  
The resulting mixed-integer program can be summarized as:
\begin{equation*}
    \begin{array}{llrr}
        \text{maximize} \, & \sum_{j=1}^M \log(p_j) E_j + \log(1-p_j)(1-E_j) \nonumber & \\
        \text{subject to} \, & \sum_{E_j \in I(C_i)} E_j - 2 K_i = \frac{1}{2}(1+C_i) & \forall i = 1,\dots,N \\
        & K_i \in \mathbb{Z}_{\geq 0} & \forall i = 1,\dots,N & \\
        & E_j \in \{0, 1\} & \forall j = 1,\dots,M.\nonumber \\
    \end{array}
\end{equation*}
To correct a given set of check measurements $\{C_i\}_{i=1\dots N}$, we solve this mixed-integer program to optimality using Gurobi, a state-of-the-art solver~\cite{gurobi}, and apply the physical correction associated with the error indices $j$ for which $E_j = 1$ in the optimal assignment.

\subsection{Belief-HUF decoder \label{subsection:bhuf}}
The original union-find (UF) decoder is a heuristic, almost-linear time surface code decoder which takes as input the \textit{decoding Tanner graph}~\cite{Delfosse2021almostlineartime}. The decoding Tanner graph has vertices corresponding to both the checks and error mechanisms of the original decoding hypergraph, and an edge between each check vertex and error mechanism vertex which flips that check. 
UF operates in two main stages: first, it associates each Tanner graph vertex with a cluster, and grows each cluster along the edges of the Tanner graph, ceasing growth on a particular cluster once it is \textit{satisfied}, meaning it contains a subset of error mechanisms which (if they occured) would generate a syndrome pattern consistent with the observed checks in that cluster. 
If two clusters meet, they merge into a single cluster. 
This process continues until each cluster is satisfied. 
In the second stage, UF applies the peeling decoder~\cite{delfosse2020lineartimemaximum} to each cluster to find an error consistent with the measured checks, and applies a correction based on the predicted errors.

\begin{figure}
    \includegraphics[width=0.5\textwidth]{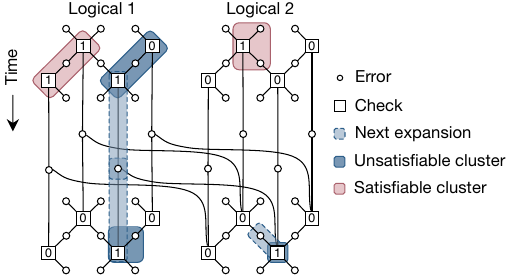}
    \caption{Hypergraph union-find decoder. 
    Clusters of vertices, each of which corresponds to a check or an error mechanism, are defined on the decoding Tanner graph. Each cluster expands along edges and merges with nearby clusters until it contains an error configuration consistent with its measured checks.  }
    \label{fig:figure_6}
\end{figure}

A key component of UF is the termination condition, which signifies that a cluster is satisfied. 
On the surface code, UF typically has two termination conditions: the total number of flipped checks in that cluster is even, or the cluster contains an error mechanism that flips only one check. 
Both of these conditions indicate that the cluster is satisfied, since errors in the bulk of the surface code only flip a pair of checks~\cite{fowler_surface_code}. 
Note that this termination condition for a cluster can be evaluated in constant time by tracking the parity of the number of flipped checks, and the existence of an error mechanism in the cluster which flips one check. 

For logical circuits with transversal entangling gates, the presence of high-order errors (such as stabilizer measurement errors near CNOTs) requires the termination condition to be generalized (see Fig.~3(a), main text).
Following Ref.~\cite{Delfosse_2014}, we replace this condition with the satisfiability of a linear system of equations, which directly checks whether a satisfying solution exists for the current cluster. 
A benefit of this approach is that solving the linear system of equations automatically provides the associated correction. 
As a trade-off, this termination condition increases the time complexity of the decoder by a polynomial overhead.

Concretely, the hypergraph union-find (\huf{}) decoder works by expanding clusters on the \textit{decoding Tanner graph}. 
As illustrated in Figure~\ref{fig:figure_6}, the decoding Tanner graph is a bipartite graph $\mathcal{G} = (\mathcal{V}, \mathcal{E})$ constructed from the original decoding hypergraph. 
The vertices are comprised of two types, $\mathcal{V} = \{C_i\}_{i = 1,\dots,M} \cup \{E_j\}_{j = 1,\dots,N}$ where each $C_i$ or $E_j$ represents a check or an error mechanism, respectively, in the original decoding hypergraph. 
The edges $\mathcal{E}$ form a subset of $\{C_i\}_{i = 1,\dots,M}\times \{E_j\}_{j=1,\dots,N}$, such that $(C_i, E_j)\in \mathcal{E}$ if and only if the edge $E_j$ is incident to $C_j$ in the original hypergraph, i.e., $E_j \in I(C_i)$.
Edges are weighted according to the error probability of its connecting error mechanism: 
\begin{equation}
    w_{E_j,C_i}=\log \Big(\frac{p_j}{1 - p_j}\Big)r_{E_j}^\varepsilon, \label{eq:uf_edge_weights}
\end{equation} 
where $r_{E_j}$ is the order of the hyperedge $E_j$ (meaning the number of vertices it connects in the original decoding hypergraph), and $p_j$ is the probability that error $E_j$ occured. $\varepsilon$ is a hyperparameter of the HUF decoder controlling how quickly clusters expand along higher-order hyperedges, which we take to be either $-1$ or $0$.
In Section~\ref{subsection:bell_noisy_syndrome_extraction}, we compare the performance of \huf{} for different settings of $\varepsilon.$

Clusters are then defined as subsets of vertices in the decoding Tanner graph. 
An edge is called a \textit{boundary edge} of a cluster if exactly one vertex in the edge is in that cluster. 
A vertex $v$ is called \textit{internal} to a cluster $S$, denoted as $v\in\bar{S}$, if all of its neighboring vertices belong to $S$. 
Given a sample of observed check measurements, a cluster is called \textit{satisfiable} if there exists a configuration of internal error vertices consistent with the given observed checks in the cluster. 
Otherwise, it is called \textit{unsatisfiable}. 

The satisfiability of a cluster can be checked by a linear system solver over the binary field $\ftwo$.  
Similar to the \mle{} algorithm in Section~\ref{subsection:mle}, the value of each error $E_j$ can be regarded as an $\ftwo$-valued variable, where $E_j = 1$ $(0)$ indicates that the error $E_j$ did (did not) occur. 
To be consistent with the observed check measurements, if the $i$th check $C_i = +1$ $(-1)$, then an even (odd) number of errors occurred  $I(C_i)$.
This condition can be equivalently stated as 
\begin{equation}
    C_i = \prod_{E_j\in I(C_i)}(-1)^{E_j}. 
\end{equation} 
If we introduce $\sigma_i$ such that $C_i = (-1)^{\sigma_i}$, then 
\begin{equation}
    \sigma_i = \sum_{E_j\in I(C_i)} E_j \quad\mod 2.
\end{equation} Consequently, $\sigma_i$ can also be regarded as a variable in $\ftwo$ that is linearly dependent on the error variables. 
In the rest this section, we will also refer to $\sigma_i$ as a check variable for convenience. The satisfaction check algorithm is described in Algorithm~\ref{alg:SC}.

\begin{algorithm}[H]
\caption{Satisfaction check for a cluster}
\label{alg:SC}
\begin{algorithmic}[1]
\Function{CheckSatisfaction}{$\mathcal{G=(V,E)},\vec{\sigma}, S$}
    \State Initialize $\vec{E}|_S \gets (0,\dots,0)\in \ftwo^{|S|}$.
    \State Let $C_S$ be all check vertices in $S$.
    \State Let $E_{\bar{S}}$ be all internal error vertices in $S$.
    \State Let $H$ be a $|C_S|\times |E_{\bar{S}}|$ matrix over $\ftwo$ such that \[H_{C_i, E_j} = 1 \text{ iff } (C_i,E_j)\in \mathcal{E}.\]
    \If{$H\vec{E}|_{\bar{S}} = \vec{\sigma}|_S$ has a solution}
        \State Let $\vec{E}|_{\bar{S}}$ be a solution and update $\vec{E}|_S$.
        \State \Return $(\mathtt{satisfiable}, \vec{E}|_S)$
    \Else
        \State \Return $(\mathtt{unsatisfiable}, \vec{E}|_S)$
    \EndIf
\EndFunction
\end{algorithmic}
\end{algorithm}

As detailed in Algorithm~\ref{alg:HUF}, after assigning the value of the measured checks \mbox{$\vec{\sigma} = (\sigma_1, \dots, \sigma_N)$}, the \huf{} decoder expands clusters on the decoding Tanner graph, prioritizing edges with lower weight $w_{E_j, C_i}$ in order to search for a satisfiable configuration of errors within a cluster. 
The edge weights, and therefore the cluster growth, depend on the probability $p_j$ of each error mechanism $E_j$, the number of checks $r_{E_j}$ flipped by $E_j$, and the hyperparameter $\varepsilon$.
Note that if one sets $\varepsilon = -1$ and applies Algorithm~\ref{alg:HUF} to a decoding graph without hyperedges of order $\geq 3$, then the expanding strategy reduces to the one defined on the split-edge graph proposed in Ref.~\cite{pattison2021improvedquantum}.

The computational runtime of \huf{} grows polynomially with the size of the decoding Tanner graph $M$.
In the worst case, one must merge all vertices into a single cluster to find a satisfiable error configuration, requiring $O(M+N) = O(M)$ total merge operations. 
By utilizing the union-find data structure, the complexity of each merge operation is only $O(\alpha (M+N))$, where $\alpha$ is the slowly-growing inverse Ackermann function (for all practical purposes, $\alpha\leq 3$)~\cite{delfosse2021unionfind}. 
Each merge also requires a satisfaction check (Algorithm~\ref{alg:SC}), which involves solving a linear system over $\mathbb{F}_2$. 
Naively, the linear system can be solved in $O(n^3)$ time using Gaussian elimination, where $n$ is the number of variables in the system, corresponding to a worst-case total runtime of 
\begin{equation}
    O(M(M^3+\alpha(M))) = O(M^4).
\end{equation}
However, in the circuits considered in this work, the linear system has constant sparsity, because each error mechanism flips a constant (in $M$) number of checks, and each check can be flipped by a constant number of error mechanisms. 
Therefore, the Wiedemann method can be used to solve the sparse linear system in $O(n^2 \log(n))$ time~\cite{wiedemann1986solvingsparse}. 
Consequently, the time complexity of the \huf{} decoder in this setting is at most 
\begin{equation}
    O(M(M^2\log(M)+\alpha(M))) = O(M^3\log(M)).
\end{equation}
Note that in practice, a merged cluster from the current round is unlikely to be updated in the next round, so the cluster update operations can be parallelized to improve the runtime.
We leave further optimization of the \huf{} runtime as an avenue for future research. 

The performance of \huf{} can be further augmented by using belief propagation (BP) before cluster expansion~\cite{belief_matching_2023}, yielding the \bhuf{} algorithm we use in the main text. 
BP takes the decoding Tanner graph and the check measurements $\vec{\sigma}$ as input and, if converged, outputs the posterior probability of each error mechanism $E_j$, given $\vec{\sigma}$. 
Although BP is not guaranteed to converge because the decoding Tanner graphs of quantum codes have short loops, it can nevertheless be used to improve the performance of \huf{}.
Following Ref.~\cite{belief_matching_2023}, we use the approximate posterior probabilities predicted by BP to update the error probabilities $p_j$ used to weight the edges of the decoding Tanner graph in Eq.~\eqref{eq:uf_edge_weights}. 
\huf{} then uses this updated decoding Tanner graph as input, expanding clusters at rates based on the updated probabilities.
We use the sum-product message update rule, and execute a constant number of update rounds for the BP algorithm used in \bhuf{}. 
We find in Section~\ref{subsection:bell_noisy_syndrome_extraction} that \bhuf{} significantly outperforms \huf{} in decoding the transversal CNOT with noisy syndrome extraction. 

\begin{algorithm}[H]
\caption{Hypergraph union-find decoder}
\label{alg:HUF}
\begin{algorithmic}[1]
\Function{HypergraphUnionFind}{$\mathcal{G},w,\vec{\sigma}$}
    \State Initialize a dictionary $f: \mathcal{E} \to \mathbb{R}$ to record the expansion state, such that $f(C_{i},E_{j}) \gets 0$
    \State Initialize a union-find data structure to maintain clusters of vertices in $\mathcal{G}$: \[
        \mathcal{C} \gets \{ \{ C_{1} \},\dots,\{ C_{M} \}, \{ E_{1} \}, \dots,\{ E_{N} \} \};
    \]
    \State Initialize a vector of unsatisfiable clusters as \[
        \mathcal{C}_{\mathrm{unsat}} \gets [ \{ C_{j} \} \mid \sigma_{j} = 1,\ j = 1,\dots, M ];
    \]
    \While{$\mathcal{C}_{\mathrm{unsat}} \neq \emptyset$}
        \State{Let $S \in \mathcal{C}_{\mathrm{unsat}}$ be the cluster with the smallest size and least recently updated}
        \State{Delete $S$ from $\mathcal{C}_{\mathrm{unsat}}$}
        \State{Let $B_{S}$ be the set of boundary edges of $S$}
        \State{Let $\Delta_{f} \gets \min_{(C_{i},E_{j})\in S} w_{C_{i}, E_{j}} - f(C_{i}, E_{j})$}
        \For{$(C_{i}, E_{j}) \in B_{S}$}
            \State $f(C_{i},E_{j}) \gets f(C_{i},E_{j}) + \Delta_{f}$ 
        \EndFor
        \For{$(C_{i},E_{j})\in B_{S}$ s.t. $f(C_{i},E_{j}) = w_{C_{i},E_{j}}$}
            \If{$C_i \in S$}
            \State Let $S'$ be the cluster that contains $E_j$
            \Else
            \State Let $S'$ be the cluster that contains $C_i$
            \EndIf
            \State Let $S \gets \Call{Merge}{\mathcal{C}, S, S'}$
        \EndFor
        \State $(R, \vec{E}|_S) \gets \Call{CheckSatisfaction}{\mathcal{G}, \vec{\sigma}, S}$
        \If{$R$ is $\mathtt{unsatisfiable}$}
            \State Add $S$ into $\mathcal{C}_{\mathrm{unsat}}$
        \EndIf
    \EndWhile
    \State \Return A global error configuration $\vec{E} = \bigcup_{S\in\mathcal{C}} \vec{E}|_S$
\EndFunction
\end{algorithmic}
\end{algorithm}

\section{Numerical simulations}
Here we describe the numerical simulations conducted to evaluate the performance of our correlated decoding algorithms. 
In general, in order to simulate a logical circuit, three components are specified: 
\begin{itemize}
    \item A \textit{physical circuit} with \textit{syndrome} measurements and \textit{logical observable} measurements.
    \item A \textit{noise model} applied to the physical circuit.
    \item A \textit{decoder}.
\end{itemize}
The logical error rate is estimated by Monte-Carlo sampling the physical errors over the chosen noise model, then generating the syndrome and logical observable measurements corresponding to the sampled set of physical errors. 
The decoders then predict which errors occurred based on the syndrome measurements, and correct the logical observable based on this prediction.
A logical error occurs if the corrected logical observable measurement differs from the true, noiseless value. 
In this work, we only simulate Clifford circuits with a noise model consisting of Pauli errors, with the exception of Section~\ref{section:transversal_ccz}. 
These simulations can be performed in polynomial time in the number of physical qubits, according to the Gottesman-Knill theorem~\cite{gottesman1997stabilizer}. 
We use the open-source Clifford simulation package, Stim~\cite{gidney2021stim}, to simulate the circuits studied in this work.
Unless otherwise stated, the error bars for the numerically computed logical error rates are the Clopper-Pearson confidence interval based on a Beta distribution with a significance level of $0.05$.
This choice corresponds to a 95\% likelihood that the true logical error rate lies between the error bars, and is comparable to significance levels used in prior works~\cite{Wu2022, bausch2023learning}.

\subsection{Surface code Bell state  with perfect syndrome measurements \label{subsection:bell_perfect_syndrome_extraction}}
Here we give additional details for the numerical simulations in Figure~2 of the main text, which study preparing a logical Bell pair using a transversal CNOT.
Using Stim~\cite{gidney2021stim}, we first noiselessly prepare two rotated surface codes in $\ket{+_L0_L}$, such that all stabilizers are initially $+1$. 
Next, we perform a noisy transversal CNOT by applying physical CNOT gates between the logical qubits. 
We insert single-physical-qubit $X$ and $Z$ errors with probability $p\fbefore$ before the CNOT and $p(1-\fbefore)$ after the CNOT. 
Finally, we measure the physical qubits in the $X$ or $Z$ basis, allowing us to determine the stabilizer measurements and logical observables in that basis.
We compare the MLE decoder described in Section~\ref{subsection:mle} against independently decoding the surface codes using the MWPM algorithm implemented in the open-source package PyMatching~\cite{higgott2021pymatchingpython, higgott2023sparse}.

As with the other logical circuits in this work, we use Stim to construct the decoding hypergraph. 
Because the codes are initialized perfectly, we define the checks simply as the stabilizers measured at the end of the circuit, as no information is gained from comparing the final stabilizer measurement with the $+1$ stabilizers during state preparation. 
These checks form the vertices of the hypergraph.
The hyperedges correspond to the possible single-physical-qubit $X$ and $Z$ errors occuring before or after the CNOT. 
As illustrated in Fig.~1(b) of the main text, physical $X$ $(Z)$ errors before the CNOT on the control (target) qubit are transferred to the other logical qubit, in the bulk flipping four checks with probability $p\fbefore$, two for each code. 
Physical $X$ $(Z)$ errors before the CNOT on the target (control) qubit are not transferred to the other qubit, as well as errors after the CNOT.
In the bulk, these errors only flip two checks on one surface code.
Because MWPM cannot handle error sources that flip more than two checks, we decompose the hyperedges that flip four checks into two existing hyperedges that flip two checks within a code. 
We perform the decomposition by setting \texttt{decompose\_errors = True} in Stim. 
This setting attempts to decompose each hyperedge that flips more than two checks into existing edges which flip one or two checks, if possible.
Because these decomposed edges already exist in the hypergraph, the error probabilities of the decomposed edges are combined with the existing error probabilities (see Ref.~\cite{gidney2021stim}).

To compare the performance of \mle{} and MWPM, we estimate the threshold $\pth$ of both algorithms at different values of $\fbefore$. 
To estimate the logical error rate for each value of $\fbefore$, physical error rate $p$, and code distance $d$, we sample at least $40000$ total samples or $6000$ samples with logical errors, whichever condition is met first. 
For each $\fbefore$, we estimate the threshold by fitting the logical error rate $P_L$ to the universal scaling hypothesis~\cite{watson2014logicalerror}, 
\begin{align}
    \log{P_L} = a+bx+cx^2,
\end{align} 
where $x = (p-\pth)^{d^{1/\nu}}$. We fit the parameters $a, b, c, \pth, \nu$ using a least-squares fit for physical error rates $p = [p_\text{th,\,guess}-0.017, p_\text{th,\,guess} + 0.027]$ and code distances $d\in \{17, 19, 21, 23, 25\}$, where $p_\text{th,\,guess}$ is an initial guess for the threshold which is close to the fitted value.
The error bars plotted in Figure~2(c) of the main text are the $2\sigma$ error bars from the fit.

\begin{table}[t]
\begin{center}
\renewcommand{\arraystretch}{1.5}
\caption{\label{table:thresholds-d-cnot-d}Threshold comparison between different decoding algorithms.}
\begin{tabular}{cccc}
\hline \hline
\textbf{Decoder} & \textbf{Hyperparameters} & \textbf{Decompose errors} & \textbf{Threshold} \\
\hline
MWPM & --- & $\mathtt{True}$ & $\simeq 0.49\%$ \\
\huf{} 1 & $\varepsilon = -1$ & $\mathtt{False}$ & $\simeq 0.31\%$ \\
\huf{} 2 & $\varepsilon = 0$ & $\mathtt{False}$ & $\simeq 0.59\%$ \\
\huf{} 3 & $\varepsilon = -1$ & $\mathtt{True}$ & $\simeq 0.53\%$ \\
\huf{} 4 & $\varepsilon = 0$ & $\mathtt{True}$ & $\simeq 0.78\%$ \\
\bhuf{} & $\varepsilon = 0, \texttt{bp\_rounds = 5}$ & $\mathtt{True}$ & $\simeq 0.95\%$ \\
\mle{} & --- & $\mathtt{False}$ & $\simeq 1.02\%$ \\
\hline \hline
\end{tabular}
\end{center}
\end{table}

\subsection{Surface code Bell state with $2d$ rounds of noisy syndrome measurements \label{subsection:bell_noisy_syndrome_extraction}}
\begin{figure*}[t]
    \includegraphics{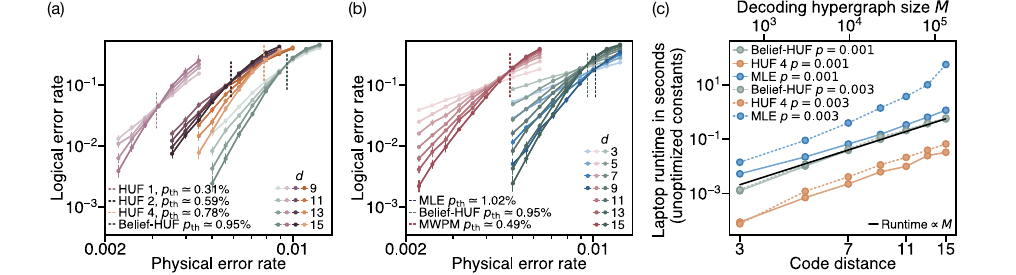}
    \caption{Surface code Bell state with $2d$ rounds of noisy syndrome extraction.
    (a)~The threshold of \bhuf{} exceeds that of HUF~1 (light purple), HUF~2 (dark purple), and HUF~4 (orange).
    (b)~\Bhuf{} and \mle{} have similar thresholds, and both algorithms outperform MWPM. 
    (c)~The laptop runtimes of \huf{}~4 and \bhuf{} grow polynomially with code distance for $p=0.001$ (solid lines) and $p=0.003$ (dotted lines), whereas the runtime of \mle{} appears to grow more rapidly for $p=0.003$. 
    The fitted laptop runtime of \bhuf{} for $p=0.001$ is approximately linear in the size of the decoding hypergraph $M$ (black).
    }
    \label{fig:figure_7}
\end{figure*}

Next, we consider using a transversal CNOT to prepare a Bell state between two rotated surface codes, with $d$ rounds of noisy syndrome extraction before and after the CNOT, and circuit-level noise. 
We compare the performance of several decoders: MWPM (using PyMatching~\cite{higgott2021pymatchingpython, higgott2023sparse}), \mle{}, \bhuf{}, and \huf{} with several hyperparameter settings.
In Table~\ref{table:thresholds-d-cnot-d} and Figure~\ref{fig:figure_7}(a) and (b), we show the approximate thresholds for each algorithm, estimated from the physical error rate at which the logical error rates of the two largest code distances intersect. 
For each algorithm, we obtain at least 10000 noisy samples per physical error rate and code distance.

To generate the circuit, we first initialize the rotated surface codes in $\ket{+_L 0_L}$. 
We initialize a code in $\ket{0_L}$ ($\ket{+_L}$) by initializing the physical qubits in $\ket{0}$  ($\ket{+}$), then measuring the $X$ and $Z$ stabilizers $d$ times. 
During each round of stabilizer measurements, we first reset the ancilla qubits used to measure the $X$ ($Z$) stabilizers in $\ket{+}$ ($\ket{0}$), then perform four CNOT gates between an ancilla qubit and its corresponding data qubits, using the same gate ordering and designation of control and target qubit as in Ref.~\cite{google_2023}, Fig.~S12. 
Finally, we measure the ancilla qubits corresponding to the $X$ ($Z$) stabilizers in the $X$ ($Z$) basis. 
After state preparation, we perform a transversal CNOT controlled on the first qubit to generate a Bell pair, followed by $d$ additional rounds of syndrome extraction, then a final projective measurement of the physical qubits in either the $X$ or $Z$ basis.

Throughout the circuit, we apply the circuit-level noise model in Ref.~\cite{belief_matching_2023} to each physical operation.
This same noise model is considered in the deep logical transversal Clifford circuits in Section~\ref{subsection:experiment-random-deep-transversal}.
Although this noise model is similar to standard noise models considered in prior theoretical results, we note that these decoding techniques have also been applied to realistic device noise models in the experiments of Ref.~\cite{Bluvstein2024}.
In this noise model, each physical two-qubit gate is followed by uniform two-qubit depolarizing noise with probability $p$. 
Qubits involved in a single-qubit gate or qubits which idle during a gate duration experience a single-physical-qubit depolarizing channel with probability $p$. 
Physical qubits initialized in the $X$~($Z$) basis are followed by a $Z$~($X$) error with probability $2p/3$, and measurements in the $X$~($Z$) basis are preceded by a $Z$~($X$) error with probability $2p/3.$

We decode using the $2d$ rounds of stabilizer measurements and the syndromes extracted from the final projective measurement. 
The checks of the decoding hypergraph are constructed from the product of each stabilizer measurement and the previous measurement of its backwards-propagated operator, if one exists. 
This means that a check comparing stabilizer measurements directly before and after the transversal CNOT can include stabilizer measurements from both the control and target logical qubit, as described in the main text. 
A check comparing two consecutive rounds of stabilizer measurements will simply be the product of those measurements.
Using these checks, the decoding hypergraph is then constructed using Stim~\cite{gidney2021stim}.
Finally, we compute the mean populations of $X_L^1 X_L^2$ and $Z_L^1 Z_L^2$ as a proxy for fidelity~\cite{entanglement_witness}, as in Section~\ref{subsection:bell_perfect_syndrome_extraction}.

\subsubsection{MWPM}
First, we discuss the decoding strategy of MWPM. 
As discussed in the main text, stabilizer measurement errors directly before the transversal CNOT generate order-three hyperedges in the decoding hypergraph. 
These hyperedges cannot be decomposed into a set of existing lower-order errors which together produce the same syndrome as the original hyperedge. 
Consequently, applying MWPM directly to the full decoding hypergraph will not yield a threshold, as MWPM cannot handle these order-three hyperedges.
For example, if these hyperedges are removed from the decoding hypergraph in order to apply MWPM, then a perfect matching for a stabilizer measurement error directly before the CNOT (an order-three hyperedge) must connect to an order-one hyperedge on the boundary of the surface code, as errors in the bulk of the surface code only flip two checks.
This procedure can erroneously apply a physical correction to the data qubits along the pairing, potentially creating a logical error. 

In order to apply MWPM, we instead decode the circuit up to the CNOT, apply the associated correction, then decode and correct the circuit after the CNOT.
To decode the first $d$ syndrome extraction rounds before the CNOT, we associate higher-order hyperedges arising from stabilizer measurement errors directly before the CNOT with a time boundary in the decoding hypergraph. 
For example, a stabilizer measurement error directly before the CNOT flips a single check before the CNOT, and two checks after the CNOT (see main text). 
This generates an order-one time boundary edge in the decoding hypergraph before the CNOT, which can be handled by MWPM. 
We apply the correction associated with the edges in the matching, which resets the check measurements before the CNOT to $+1$. 
As a result of applying this correction, if the matching is paired to a stabilizer measurement error time boundary near the CNOT, we also flip the other checks after the CNOT that are affected by that error.
We then decode and correct the final $d$ rounds of syndrome extraction normally using MWPM.

To construct the decoding hypergraphs used to decode before and after the CNOT, we take the decoding hypergraph of the full circuit generated by Stim~\cite{gidney2021stim}, and split it into two halves. 
The first half includes only checks whose stabilizer measurements occurred before the CNOT. 
Its hyperedges include all error mechanisms which only flip checks before the CNOT, as well as hyperedges connecting checks from both halves (including the order-three hyperedges from stabilizer measurement errors).
These hyperedges on the boundary of the halves are filtered to only include their connections to checks before the CNOT. 
This filtering process may result in error sources which, when restricted to the checks before the CNOT, flip identical sets of checks. 
We merge these duplicate error sources into a single hyperedge with a combined probability $\sum_i \big(\frac{p_i}{1-p_i} \prod_j (1-p_j) \big)$ where $p_i$ is the probability of each error source $i$ in a group of duplicates. 
We use this modified hypergraph to decode before the CNOT, using the \texttt{decompose\_errors=True} hyperparameter in Stim.
To construct the decoding hypergraph after the CNOT, we only keep error mechanisms which flip checks solely after the CNOT.
By using this hypergraph splitting strategy, MWPM achieves a threshold of $\pth \simeq 0.49\%$~(Table~\ref{table:thresholds-d-cnot-d}). 

Note that this decoding approach requires $d$ syndrome extraction rounds before the CNOT to be fault-tolerant to stabilizer measurement errors before the CNOT.
As a result, this procedure is not fault-tolerant for logical algorithms with $\simeq 1$ syndrome extraction round between transversal CNOTs.

\subsubsection{Correlated decoding algorithms}
In contrast to MWPM, the correlated decoding algorithms we consider can utilize general decoding hypergraphs.
Therefore, they can naturally handle the higher-order hyperedges arising from stabilizer measurement errors directly before a transversal CNOT. 
Here we benchmark four variants of \huf{}, which have different settings of the hyperparameter $\varepsilon$, which affects how quickly clusters grow along higher-order hyperedges  (see Section~\ref{subsection:bhuf}).
We also explore different settings of the \texttt{decompose\_errors} parameter in Stim, which describes whether or not high-order hyperedges are decomposed into existing low-order hyperedges, resulting in an input decoding hypergraph with only essential hyperedges~\cite{gidney2021stim}.
The thresholds of \bhuf{}, \mle{}, and most variants of HUF exceed that of the MWPM (Table~\ref{table:thresholds-d-cnot-d}).
Furthermore, the threshold of \bhuf{} exceeds that of the \huf{} variants, and is comparable to that of \mle{} (Fig.~\ref{fig:figure_7}(a) and (b)).

We also observe that the hyperparameter settings for \huf{} have a significant effect on performance (Table~\ref{table:thresholds-d-cnot-d}). 
For example, \huf{}~3 and \huf{}~4 gain a remarkable improvement over \huf{}~1 and \huf{}~2 by setting $\mathtt{decompose\_error = True}$.
This is because although \huf{} can in principle handle higher-order hyperedges, decomposing the hyperedges restricts the rate of cluster expansion, as the size of a cluster increases more rapidly when expanding along high-order hyperedges compared to low-order ones. 
Concretely, as described in Section~\ref{subsection:bhuf}, an order-$k$ hyperedge in the decoding hypergraph corresponds to $k$ equally-weighted edges in the decoding Tanner graph.
Therefore, the size of the cluster can increase by at most $k-1$ during one expansion step along this error mechanism.
Excessively large clusters can result in decoding failures, as they involve solving a global linear system to find a satisfiable error, with no preference for choosing a high-probability error.
Setting $\varepsilon = 0$ also decreases the priority by which the decoder expands clusters along hyperedges, further improving the performace of HUF~4 over HUF~3. 
Finally, the performance is further enhanced by introducing belief propagation with five rounds of belief propagation (\texttt{bp\_rounds = 5}), yielding the \bhuf{} algorithm. 
Note that both the transversal CNOT thresholds of \huf{} ($0.78\%$) and \bhuf{} ($0.95\%$) with the setting of $\mathtt{decompose\_errors = True}$ and $\varepsilon = 0$ match the thresholds of UF ($0.795(1)\%$) and belief-find ($0.937(2)\%$), respectively, in the surface code memory simulations reported in Ref.~\cite{belief_matching_2023}. 
This confirms that the correlated decoders perform optimally compared to their uncorrelated versions, as we expect the memory threshold to be an upper bound on the threshold for the circuit we study with $2d$ rounds of syndrome extraction and a transversal CNOT.

Finally, we note that although the worst-case runtime of \huf{} derived in Section~\ref{subsection:bhuf} is $O(M^3 \log(M))$, where $M$ is the size of the decoding hypergraph, in practice its runtime scaling is more favorable. 
In Figure~\ref{fig:figure_7}(c), we plot the computational runtime per shot on a laptop for \huf{}, \bhuf{}, and \mle{} for the physical error rates $p=0.001$ and $p=0.003$.
A least-squares fit to the data at $p=0.001$ to the functional form $y=ax^b$ yields a fit to the exponent parameter of $b=1.00(3)$.
Therefore, the runtime of \huf{} is linear in $M$ at this physical error rate.
We emphasize that the computational runtime of \bhuf{} is not optimized for constant factors, and we anticipate that it can be substantially improved. 
We leave further optimization of the computational runtime as a subject for future research. 

\subsection{Deep logical transversal Clifford circuits \label{subsection:experiment-random-deep-transversal}}
The deep logical Clifford circuits we study consist of four rotated surface code qubits and 32 layers of transversal gates, each consisting of single-qubit logic gates drawn from $\{H_L, X_L, Y_L, Z_L\}$, followed by a random pairing of transversal CNOTs with randomized designation of control and target. 
The transversal Pauli $X_L, Y_L,$ and $Z_L$ gates are performed by applying physical $X, Y,$ and $Z$ gates to all physical qubits within a code, respectively. 
The transversal Hadamard $H_L$ gates are performed by applying $H$ gates to all the qubits within a code, then noiselessly rotating the physical qubits by 90 degrees. 
After each layer of transversal CNOT gates, we perform $\nrounds$ rounds of syndrome extraction, using the same circuit as in Section~\ref{subsection:bell_noisy_syndrome_extraction}. 
Throughout the circuit, we apply the circuit-level noise model in Ref.~\cite{belief_matching_2023} (see  Section~\ref{subsection:bell_noisy_syndrome_extraction}), except during state preparation, the final round of stabilizer measurements in the last transversal gate layer, the logical stabilizer measurements, and code rotation during transversal $H_L$ gates, which are noiseless. 

Figure~3(c) in the main text shows the logical error rate per round $\plogicalround$ as a function of the physical error rate $p$. The logical error rate per round is given by the formula
\begin{equation}
    \plogicalround = \plogicalmax\big[1 - (1 - \plogical/\plogicalmax)^{1/(32\nrounds)}\big],\label{eq:plogicalround}
\end{equation}
where \mbox{$\plogicalmax = 1-\frac{1}{2^4} = \frac{15}{16}$} is the error rate of the maximally mixed logical state. 
This formula can be derived by treating the process as a Markov chain, assuming that $\plogicalround$ is a constant over different rounds, and that the four logical stabilizers each have an equal probability of flipping per round. 
If the input state to a round is the correct logical state, there is a probability of $1-\plogicalround$ of obtaining the correct output state. 
If the input state is an incorrect logical state, then the output state can be flipped back to the correct logical state with a probability of ${\plogicalround}/15$. 
Note that when $\plogicalround$ equals $0$ $(\plogicalmax)$, $\plogical$ achieves its minimum (maximum).

\begin{figure}[t] 
    \includegraphics{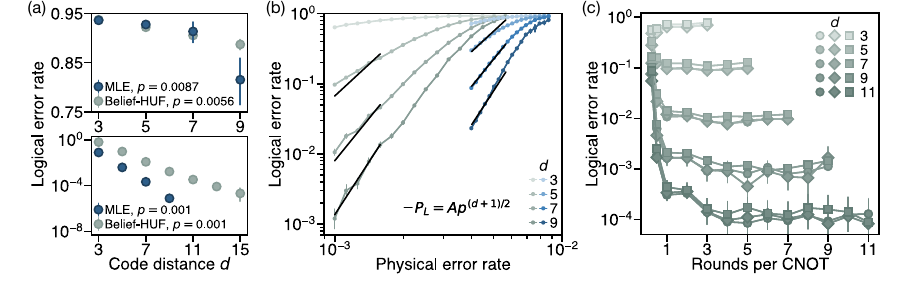}
    \caption{Benchmarking deep logical circuits.
    (a)~For the depth 32 logical circuit studied in the main text with $\nrounds = 1$ round of syndrome extraction per CNOT, the logical error rate decreases monotonically with code distance at physical error rates of $p = 0.87\%$ for \mle{} and $p = 0.56\%$ for \bhuf{} (top).
    The same trend persists at a lower physical error rate of $p = 0.1\%$ (bottom).
    (a)~We plot the logical error rate $P_L$ as a function of the physical error rate $p$ for the same circuit, and fit the coefficient $A$ in the functional form $P_L = A p^{(d+1)/2}$ to the data for the five smallest physical error rates studied (black lines).
    The quality of the fit suggests that \mle{} achieves the full code distance.
    (c)~The average logical error rates over ten deep circuits with $32$ (circles) and $40$ (squares) layers of gates are similar as a function of $\nrounds$.
    Furthermore, the depth $32$ instance studied in the main text (diamonds) has similar performance to the average over the depth $32$ circuits.
    }
    \label{fig:figure_8}
\end{figure}

Although our numerical results in Figure~3(c) of the main text are consistent with the existence of a threshold when $\nrounds = 1$, we cannot fit the exact threshold to the universal scaling hypothesis~\cite{watson2014logicalerror} because \mbox{$\plogical/\plogicalmax \to 1$} at physical error rates near the threshold. 
Nevertheless, here we obtain a lower bound on the thresholds for \bhuf{} and \mle{} by identifying the largest physical error rate consistent with a monotonic decrease in logical error rate per round with code distance. 
Figure~\ref{fig:figure_8}(a) (top) demonstrates that the thresholds of \bhuf{} and \mle{} are bounded by $\pth \geq 0.56\%$ and  $\pth \geq 0.87\%$, respectively.
Figure~\ref{fig:figure_8}(a) (bottom) additionally shows that the logical error rate decreases exponentially for a larger range of code distances, at a lower physical error rate of $p=0.1\%$. 

\begin{table}[b]
\begin{center}
\renewcommand{\arraystretch}{1.5}
\caption{\label{table:fit_parameters_spacetime_cost}Fit parameters for the heuristic scaling formula.}
\begin{tabular}{>{\centering}m{7em}>{\centering}m{7em}>{\centering\arraybackslash}m{7em}}
\hline \hline
\textbf{Parameter} & \textbf{\mle{}} & \textbf{\bhuf{}} \\
\colrule
$C$ & 1.13(21) & 0.464(65) \\
$\alpha_{\frac{1}{8}}$ & 0.097(7) & --- \\
$\alpha_{\frac{1}{4}}$ & 0.070(5) & 0.458(19) \\
$\alpha_{\frac{1}{2}}$ & 0.055(4) & 0.240(10) \\
$\alpha_1$ & 0.049(3) & 0.170(7) \\
$\alpha_2$ & 0.039(3) & 0.146(6) \\
$\alpha_3$ & 0.039(3) & 0.120(5) \\
$\alpha_4$ & 0.044(2) & 0.109(4) \\
$\alpha_5$ & 0.041(2) & 0.102(4) \\
$\alpha_6$ & --- & 0.103(4) \\
$\alpha_7$ & --- & 0.097(4) \\
\hline \hline
\end{tabular}
\end{center}
\end{table}

In Figure~\ref{fig:figure_8}(b), we show the logical error rate $P_L$ as a function of the physical error rate $p$. 
We fit the coefficient $A$ in the functional form $P_L = A  p^{(d+1)/2}$ for the smallest five physical error rates $p$ studied for both \bhuf{} and \mle{} (black lines).
This quality of the fits suggests that \mle{} achieves the full code distance, with exponential suppression of $P_L$ in $(d+1)/2$. 
In contrast, the results for \bhuf{} are less clear, and merit future research. 

Next, we investigate whether the spacetime cost of the logical circuit needed to reach a fixed logical error rate can be reduced by optimizing $\nrounds$. 
To do so, we collect data at different values of $\nrounds$ and $d$, displayed in Fig.~4(a) in the main text.
The error bars in Figure~4(a), as with the rest of the text, correspond to the Clopper-Pearson confidence interval based on a Beta distribution with a significance level of $0.05$.
We use this data to fit the parameters $C$ and $\alpha_{\nrounds}$ in the heuristic formula,
\begin{equation}
    P_L(\nrounds, d) = 32 C \nrounds (\alpha_{\nrounds})^{\frac{d+1}{2}}, \label{eq:heuristic_pl}
\end{equation}
at different $\nrounds$. 
This formula comes from applying the standard formula for logical error rate as a function of code distance~\cite{fowler_surface_code}, modified so that the prefactor, $32 C \nrounds$, grows linearly with the circuit depth $32 \nrounds$, similar to the numerical simulations in Ref.~\cite{beverland_2021}.
The fitted parameters for \bhuf{} and \mle{}, obtained from a least-squares fit to the logarithm of Eq.~\eqref{eq:heuristic_pl}, are listed in Table~\ref{table:fit_parameters_spacetime_cost}. 
We only fit the exponent base $\alpha_{\nrounds}$ at values of $\nrounds$ for which we have data for at least three different code distances, and for which the decoder is above the threshold at $p=0.001$ (the latter condition is not fulfilled for \bhuf{} at $\nrounds = 1/8$).
Based on the fitted parameters, we can solve for the potentially fractional code distance needed to reach a target logical error rate of $P_L = 10^{-6}$ for each value of $\nrounds$. 
This gives us the extrapolated spacetime cost, expressed as $\nrounds d^2$, plotted in Fig.~4(b) in the main text. 
Note that the fitted exponent base $\alpha_{\nrounds}$ should not be interpreted as an accurate estimate of the ratio $p/\pth$, as previously observed in the literature~\cite{Gidney2021faulttolerant, fowler_surface_code}. 
Additionally, we note that the fitted parameters may be subject to a small systematic errors due to the finite range of code distances studied. 
However, the resulting estimation error in the spacetime cost is likely small, as the target logical error rate $P_L = 10^{-6}$ is not significantly smaller than the minimum measured logical error rates $P_L\simeq 10^{-4}$.

Finally, we study whether our results are sensitive to finite size effects.
In Figure~\ref{fig:figure_8}(c), we compute the logical error rate as a function of $\nrounds$, averaged over ten randomly generated circuits with depths 32 and 40 (circles and squares, respectively). 
The results for both circuit depths are qualitatively similar, and quantitatively differ by an overall constant factor.
This suggests that our results are not subject to finite-size effects due to the depth of the circuit. 
Furthermore, in Figure~\ref{fig:figure_8}(c) we observe that the data for different random circuits at depth 32 is quantitatively similar to the particular depth 32 circuit studied in the main text (diamonds), indicating that our results are representative of other randomly generated circuits.

\subsection{Transversal CCZ with perfect syndrome measurements\label{section:transversal_ccz}}
Here we study the performance of the transversal CCZ between three quantum Hamming codes, each comprised of fifteen physical qubits encoding seven logical qubits, with code distance three. 
This code is self-dual, meaning that the $X$ and $Z$ stabilizers are supported on the same physical qubits. 
The check matrix for each of the $X$ and $Z$ stabilizers is given by
\setcounter{MaxMatrixCols}{20}
\begin{equation}
    H = 
\begin{pmatrix}
0&0&0&0&0&0&0&1&1&1&1&1&1&1&1\\
0&0&0&1&1&1&1&0&0&0&0&1&1&1&1\\
0&1&1&0&0&1&1&0&0&1&1&0&0&1&1\\
1&0&1&0&1&0&1&0&1&0&1&0&1&0&1\\
\end{pmatrix},
\end{equation}
where $H_{ij} = 1$ if qubit $j$ is in the support of check $i$, and $0$ otherwise. Each code block has seven logical qubits, six of which are gauge qubits, and one of which is used for the transversal CCZ. 
We work in the basis where the representation of the latter is transversal ($X_L = \bigotimes_{i=1}^{15} X_i$, $Z_L = \bigotimes_{i=1}^{15} Z_i$)~\cite{paetznick_ccz_2013}. 

The circuit we simulate prepares the three codes perfectly in an initial product state in the logical $X$ or $Z$ basis, then applies a single-physical-qubit depolarizing channel with probability $p$ to the physical qubits, followed by a transversal CCZ. 
Because the transversal CCZ gate is non-Clifford, it cannot directly be simulated with a Clifford circuit via the Gottesman-Knill theorem~\cite{gottesman1997stabilizer}.
Nevertheless, for particular initial states, the circuit can be efficiently simulated using a Clifford circuit, enabling us to probe the performance of the transversal CCZ on those initial states.

In particular, we efficiently simulate the circuit for initial product states for which the CCZ generates a known logical stabilizer state, $\ket{\psi_L}$. 
Some example initial states fulfilling this condition are $\ket{1_L 1_L 1_L}, \ket{1_L 1_L +_L},$ and $\ket{1_L +_L +_L}$, which after a CCZ generate the states $\ket{1_L 1_L 1_L}, \ket{1_L 1_L -_L},$ and \mbox{$\ket{1_L}(\ket{0_L +_L} + \ket{1_L -_L}) / \sqrt{2}$}, respectively. 
Notice that the output logical stabilizer state $\ket{\psi_L}$ can be efficiently constructed from a Clifford circuit using the Aaronson and Gottesman algorithm~\cite{Aaronson_2004}.
To simulate the noisy circuit efficiently, we therefore observe that introducing Pauli noise before the CCZ gate is equivalent to perfectly preparing $\ket{\psi_L}$, then subsequently applying the Clifford error obtained by propagating the Pauli error before the CCZ through the CCZ gate. 

Therefore, we simulate the circuit by first generating $\ket{\psi_L}$ efficiently, then add the Clifford noise onto $\ket{\psi_L}$.  
To sample the Clifford error, we Monte-Carlo sample a random Pauli error before the CCZ drawn from a single-physical-qubit depolarizing channel, then propagate the sampled Pauli errors through the CCZ operation. 
In general, $Z$ errors commute through the CCZ, and $X$ ($Y$) errors propagate to an $X$ ($Y$) error on the same qubit and a CZ error on the other two qubits in the CCZ. 
We then apply this propagated Clifford error onto the perfect state $\ket{\psi_L}$.
Because the error itself is a Clifford gate, we can simulate the entire circuit efficiently~\cite{gottesman1997stabilizer} by repetitively sampling different error patterns.

We complete the circuit by noiselessly measuring the $X$ and $Z$ stabilizers and the logical stabilizer operators. 
The logical error rate for a given initial state is then computed from the corrected logical stabilizers~\cite{gottesman1998heisenberg}, averaged over 20000 shots.
To compute the total logical error rate, we average the logical error rate over five possible initial logical states: $\ket{0_L0_L0_L}, \ket{0_L0_L+_L}, \ket{0_L+_L+_L}, \ket{1_L1_L+_L}, $ and $\ket{1_L+_L+_L}$. 
This choice comes from first symmetrizing over the three logical qubits, then using the fact that the $\ket{0_L}/\ket{1_L}$ and $\ket{+_L}/\ket{-_L}$
states will behave identically under noise. 

To decode, we construct an approximate decoding hypergraph for the logical circuit.
The decoding hypergraph is approximate because errors before the CCZ can propagate to a Clifford error after the CCZ~\cite{scruby_nonpauli_2022}, which can equivalently be written as a superposition of Pauli errors. 
Therefore, each error mechanism does not necessarily deterministically flip a particular set of checks. 
As an approximation, we therefore construct the decoding hypergraph to include a hyperedge for every Pauli error in the superposition. 
Concretely, for each group of three physical qubits undergoing a CCZ, we compute the input distribution of Pauli errors, e.g. the error $XIZ$ occurs with probability $(p/3)^2(1-p)$. 
We then propagate each input Pauli error through the circuit, giving us a superposition of Pauli errors, e.g. $XIZ$ goes to $\frac{1}{2}(XIZ + XII + XZZ - XIZ)$. 
We add a hyperedge to the decoding hypergraph corresponding to each term in the superposition. 
For example, the above error will generate hyperedges corresponding to the errors $XIZ, XII, XZZ,$ and $XIZ$.
To approximate the probability of each output error, we take the probability of the error before the CCZ, and normalize by the number of outputs (e.g., each of the above errors is associated with a probability of $(p/3)^2(1-p)/4$). 
If two different Pauli error sources propagate to the same error (e.g., the errors $XII$ and $XIZ$ will both propagate to a superposition of Pauli errors which includes $XZI$), then we sum the resulting error probabilities in the decoding hypergraph.

To compare correlated decoding against independently decoding the logical qubits, we construct two different decoding hypergraphs: one which includes all error sources, as described above, and one which includes only error sources acting with a single logical qubit. 
For correlated decoding, we use the \mle{} algorithm (Section~\ref{subsection:mle}) with the former decoding hypergraph as input.
To independently decode the qubits, we use \mle{} with the latter decoding hypergraph as input.

\end{document}